\documentclass[11pt, a4paper]{article}
\usepackage{jheppub}
\usepackage{bm} \usepackage{mathptmx}\usepackage{amsfonts}

\usepackage{latexsym}
\newcommand{\nn}{\nonumber}
\newcommand{\be}{\begin{equation}}
\newcommand{\ee}{\end{equation}}
\newcommand{\ba}{\begin{eqnarray}}
\newcommand{\ea}{\end{eqnarray}}
\newcommand{\bal}{\begin{align}}
\newcommand{\eal}{\end{align}}
\newcommand{\lb}{\label}

\newcommand{\bw}{\begin{widetext}}
\newcommand{\ew}{\end{widetext}}

\def\6{_{6}}
\def\5{_{5}}
\def\4{_{4}}

\def\A{{\cal A}}
\def\F{{\cal F}}

\def\ol{\overline}

\def\M{{\cal M}} \def\X{\bar{X}} \def\Y{\bar{Y}} \def\Z{\bar{Z}}
\newcommand{\cM}{{\cal M}}

\newcommand{\hmu}{\hat{\mu}}
\newcommand{\hnu}{\hat{\nu}}
\newcommand{\hrho}{\hat{\rho}}
\newcommand{\hlambda}{\hat{\lambda}}

\newcommand{\hsigma}{\hat{\sigma}}
\newcommand{\htau}{\hat{\tau}}

\renewcommand{\theequation}{\arabic{section}.\arabic{equation}}

\title{New coset matrix for \boldmath $D=6$ self-dual supergravity}

\author[a]
{G\'erard Cl\'ement}
\author[b]{Dmitri V. Gal'tsov}
 \affiliation[a]{
Laboratoire de  Physique Th\'eorique LAPTH (CNRS),
\\
B.P.110, F-74941 Annecy-le-Vieux cedex, France}
  \affiliation[b]{Department of Theoretical
Physics, Moscow State University,\\ 119899, Moscow, Russia}
\emailAdd{gclement@lapp.in2p3.fr} \emailAdd{galtsov@phys.msu.ru}

\abstract  {Toroidal  reduction of minimal six-dimensional
supergravity, minimal five-dimensional supergravity and
four-dimensional Einstein-Maxwell gravity to three dimensions gives
rise to a sequence of cosets
 $O(4,3)/(O(4)\times O(3))\supset G_{2(2)}/(SU(2)\times
SU(2))\supset SU(2,1)/S(U(2)\times U(1))$ which are invariant
subspaces of each other. The known matrix representations of these
cosets, however, are not suitable to realize these embeddings which
could be useful for solution generation. We construct a new
representation of the largest coset in terms of $7\times 7$ real
symmetric matrices and show how to select invariant subspaces
corresponding to lower cosets by  algebraic constraints. The new
matrix representative may be also directly applied to minimal
five-dimensional supergravity. Due to full $O(4,3)$ covariance it is
simpler than the one derived by us previously for the coset
$G_{2(2)}/(SU(2)\times SU(2))$.}
  \arxivnumber{1301.5084}
\toccontinuoustrue
\begin{document}
%% REVTEX4
%\pacs{04.20.Jb, 04.50.+h, 04.65.+e}
\maketitle

\section{Introduction}
The remarkable  sequence of groups $O(7)\supset G_2\supset SU(3)$
attracted attention in  particle physics long ago. In a seminal
paper Gunaydin and G\"ursey \cite{GG} have given an extensive
discussion of their properties, representations and applications to
model building. Within the Lie algebra of $O(7)$, the subalgebras
$G_2$ and $SU(3)$ form rather sophisticated closed structures which
were explicitly given in \cite{GG} in terms of rotation generators
of $O(7)$.

 The maximally non-compact forms of
the same groups $O(4,3)\supset G_{2(2)}\supset SU(2,1)$ play an
important role in the gravity/supergravity context
\cite{Gunaydin:2007qq}. These group are hidden symmetries of
six-dimensional minimal supergravity (MSG6) \cite{Cremmer:1999du},
five-dimensional minimal supergravity (MSG5)
\cite{mizoh,Bouchareb:2007ax} and four-dimensional Einstein-Maxwell
(EM4) \cite{Ernst:1967wx,Kinnersley:1977pg} (super)gravity
respectively, which are manifest as isometries of the target spaces
of sigma models arising in their toroidal compactification to three
dimensions
\cite{Breitenlohner:1987dg,Maison:2000fj,Clement:2008qx,Galtsov:2008zz}.
More precisely, the compactified theories are gravity coupled scalar
sigma models on the coset spaces $O(4,3)/(O(4)\times O(3)),\;
G_{2(2)}/(SU(2)\times SU(2)),\;SU(2,1)/S(U(2)\times U(1))$
  if the compactification tori are purely space-like,
and $O(4,3)/(O(2,2)\times O(1,2)),\; G_{2(2)}/(SL(2,R)\times
SL(2,R)),\;SU(2,1)/S(U(1,1)\times U(1))$ if one of the reduced
dimensions is time. The last coset has been known for a long time as
the manifold where the famous Ernst-Kinnersley-Mazur
\cite{Ernst:1967wx,Kinnersley:1977pg,Mazur:1983vi} symmetry
operates. Its natural matrix representation is given in terms of
$3\times 3$ (pseudo)unitary matrices. The $G_{2(2)}$ coset was
extensively explored recently as a tool for solution generation
\cite{Bouchareb:2007ax,TYM,Compere09,Compere10} in MSG5. Fruitful
for this purpose is the novel $7\times 7$ matrix representation
 \cite{Bouchareb:2007ax,5to3} essentially related to the matrix
representation of $G_2$ given by Gunaydin and G\"ursey \cite{GG}.
The coset $O(4,3)/(O(4)\times O(3))$ constitutes a particular case
of the Hassan-Sen-Maharana-Schwarz (HSMS) cosets
$O(n+p,n)/(O(n+p)\times O(n))$ arising in toroidal compactification
of heterotic string effective theory, its matrix representation was
given in \cite{Hassan:1991mq,Maharana:1992my,Sen:1994fa}. In the
case of $O(4,3)$ theory it is also realized in terms of $7\times 7$
matrices. This representation, however, is rather complicated and
not convenient to make contact with the sequence of  subspaces
$G_{2(2)}/(SU(2)\times SU(2))$ and $SU(2,1)/S(U(2)\times U(1))$
which can be useful in relating solutions of EM4, MSG5 and MSG6
theories between themselves.

 The purpose of the present paper is to construct a new
matrix representative of the coset $O(4,3)/(O(4)\times O(3))$ which
allows for simple truncation to subspaces corresponding to MSG5 and
EM4 theories. This is based again on the $7\times 7$ representation,
but with different parametrization of moduli. The new matrix  is
much simpler than the corresponding HSMS matrix and can be truncated
to lower cosets by imposition of purely algebraic constraints. Our
derivation is based on the direct toroidal reduction of MSG6 to
three dimensions, explicit determination of target space isometry
generators and subsequent exponentiation of the Borel subalgebra. We
then extract the generators of the $G_{2(2)}$ and $SU(2,1)$
subgroups of $O(4,3)$ and derive algebraic constraints selecting the
corresponding invariant subspaces of the coset $O(4,3)/(O(4)\times
O(3))$ on which they act transitively.

\section{$D=6$ minimal supergravity}

The bosonic action of six-dimensional minimal supergravity contains
the metric and self-dual three-form field
 \be\lb{MSG6} S_{MSG6} =
\int\left(\hat{R}  - \frac1{12}
\hat{G}_{\hmu\hnu\hlambda}\hat{G}^{\hmu\hnu\hlambda}\right)\sqrt{-\hat{g}}d^6x\,,
 \ee
where $\hat{G}_{\hmu\hnu\hlambda} \equiv
3\hat{C}_{[\hmu\hnu,\hlambda]}$, with subsidiary condition
 \be\label{selfdu}
  \hat{G}_{\hmu\hnu\hlambda} = \frac16\sqrt{-\hat{g}}
\epsilon_{\hmu\hnu\hlambda\hrho\hsigma\htau}\hat{G}^{\hrho\hsigma\htau}\,,
 \ee
which has to be imposed after variation of the action.\footnote{As
in some other supergravity actions involving self-dual form fields,
the quadratic action of the type (\ref{MSG6}) does not imply the
self-duality condition (\ref{selfdu}), moreover it is zero, if
self-duality is imposed in the action itself. One needs extra fields
to construct a consistent action for chiral forms in a
Lorentz-covariant way. We thank Dmitri Sorokin for drawing our
attention to the references \cite{sorokin} where such an action for
$D=6$ minimal supergravity was presented. Here we deal with
classical equations of motion, so it will be sufficient to impose
the condition (\ref{selfdu}) by hand after variation is performed.
The dimensional reduction of the full action \cite{sorokin} is more
involved, but this does not change the results on the classical
level.} The action (\ref{MSG6}) is a lowest-dimensional member of
the even-dimensional sequence of actions containing self-dual form
fields, the largest representative of which is the IIB
ten-dimensional supergravity.

Somewhat unexpectedly, this action, being compactified on a circle,
turns out to be non-locally dual to the truncated five-dimensional
heterotic string effective action
\cite{Hassan:1991mq,Maharana:1992my,Sen:1994fa} which belongs to
another sequence of the string actions. This can be hinted from the
fact that the $D=5$ heterotic string effective action  truncated to
the one-vector case gives rise to the $D=3$ $O(4,3)/(O(4)\times
O(3))$ coset theory  (a particular case of the Sen's  coset
$O(d+1,d+1+p)/(O(d+1)\times O(d+1+p))$ where $d $ is the number of
compactified dimensions and $p$ is the number of vector fields in
the initial dimension \cite{Sen:1994wr}). Meanwhile the generic
oxidation of the $O(4,3)/(O(4)\times O(3))$ coset has apart from the
regular oxidation point $D=5$ (which is the above heterotic
effective action) also an anomalous {\em six-dimensional} oxidation
point \cite{Cremmer:1999du} which is just minimal $D=6$ self-dual
supergravity. This leads to a non-local duality between the two
theories which can be made explicit as follows.

Denoting the coordinates $x^{\hmu}= (x^\mu, z)$ and assuming
existence of the Killing vector $\partial_z$, we decompose the
metric and the two-form potential as
 \ba\lb{metric65}
ds\6^2 &=& e^{2\alpha\phi}g_{\mu\nu}dx^{\mu}dx^{\nu} +
e^{-6\alpha\phi} (dz+\A_{\mu}dx^{\mu})^2\,, \\ \lb{form65} \hat{C}
&=& \frac12B_{\mu\nu}dx^{\mu}\wedge dx^{\nu} + \frac1{\sqrt{2}}
A_{\nu} dz\wedge dx^{\nu}\,,
 \ea
where $\alpha^2=1/24$. The field equations are then equivalent to
those derived from the five-dimensional action \be\lb{O435} S_5=
\int\left(R - \frac12(\partial\phi)^2 -
\frac14e^{4\alpha\phi}F_{\mu\nu}F^{\mu\nu}
-\frac1{12}e^{8\alpha\phi}H_{\mu\nu\lambda}H^{\mu\nu\lambda}\right)\sqrt{-g\5}d^5x\,,
 \ee
 with \be\lb{56rel} \F_{\mu\nu} \equiv 2\A_{[\nu,\mu]}\,, \quad H^{\mu\nu\rho}
\equiv -\frac1{2\sqrt{-g_5}}  e^{-8\alpha\phi}
\epsilon^{\mu\nu\rho\sigma\tau}\F_{\sigma\tau}  \,,\quad
H_{\mu\nu\rho} = 3(B_{[\mu\nu,\rho]} + \frac12
F_{[\mu\nu}A_{\rho]})\,. \ee This is a heterotic string type
effective action \cite{Sen:1994fa,Sen:1994wr} with one vector and
one antisymmetric second rank tensor fields. Note that the Maxwell
field $F_{\mu\nu}$ in this action originates from the
six-dimensional three-form, while the five-dimensional three-form
$H_{\mu\nu\rho}$ is obtained by dualisation of the Kaluza-Klein
two-form. Therefore the relation between the six and
five-dimensional metrics and matter fields is non-local.

Due to this duality one can reduce the six-dimensional action
(\ref{MSG6}) (which is the subject of the present paper) to three
dimensions along two different compactification schemes. The first
consists in using the well-studied compactification of the
corresponding five-dimensional heterotic string action (\ref{O435})
along the lines of \cite{Sen:1994fa,Sen:1994wr}. The second,
suggested in the present paper, consists in  direct compactification
of the initial six-dimensional action (\ref{MSG6}) on a three-torus.

The first way which we briefly sketch here  gives Sen's type
representation for the coset matrix \cite{Sen:1994wr}. Splitting the
coordinates as $z^a,\;x^\mu= x^i,\; a =1,2,\;   i = 1,2,3 $, with
$z^a$ along the compactified dimensions, we parameterize the metric
and the matter fields as
 \ba
&&d {s}^2  =  \lambda_{ab}(dz^a + A_i^adx^i)(dz^b + A_j^bdx^j) +
\tau^{-1}
h_{ij}dx^idx^j\,, \quad  \tau = -\det\lambda  \,, \nonumber\\
 && A_\mu dx^\mu  =  \psi_a(dz^a+A_i^adx^i) -
A_i^5 dx^i\,,
\\
 &&B_{\mu\nu}dx^\mu\wedge dx^\nu=  B_{ab}(dz^a+A_i^adx^i)\wedge(dz^b+A_j^bdx^j)
 + (A_{i(a+2)} - \frac12\psi_aA_i^5 )(dz^a\wedge dx^i - dx^i\wedge dz^a) \nonumber\\
 && + (B_{ij} + A_{[i}^aA_{j](a+2)})dx^i\wedge dx^j\,. \nonumber
 \ea
The three-dimensional reduced action can be presented in terms of
the matrix sigma model
 \be
S_3=\int\left\{ R_3(h)-\frac18 {\rm Tr}
\left[\left(\partial_i\cM\right)
\cM^{-1}\left(\partial_j\cM\right)\cM^{-1}\right]h^{ij}\right\}\sqrt{h}d^3x\,.
 \ee
According to \cite{Sen:1994wr}, the coset matrix $\cM$ is
constructed in three steps: first one defines of the matrix $M$ of
non-dualized moduli, then dualisation of three-dimensional vectors
to scalar potentials is performed, and finally the matrix $\cM$ is
constructed in terms of $M$ and the new scalars. To built the moduli
matrix $M$ one arranges the five vector fields in a column matrix
$A_i^{A}$ ($A= 1,...,5$), \be A_i^A = (A_i^a,A_{i(a+2)},A_i^5 )\,,
\ee with the field strengths \be F_{ij}^A \equiv
2\partial_{[i}A_{j]}^A\,, \quad H_{ijk} =
3(\partial_{[i}B_{jk]}+\frac12 A_{[i}^AL_{AB}F_{jk]}^B)\,,
 \ee
where $L$ is the $5\times5$ matrix written in block form
 \be L =
\left(\begin{array}{ccc} 0 & 1 & 0
\\ 1 & 0 & 0 \\ 0 & 0 & -1
\end{array} \right)\,.
 \ee
The 2-form $B_{ij}$ is actually fixed by the gauge condition
 \be
H_{ijk} = 0\,. \lb{H0}
 \ee
The $5\times5$ moduli  matrix $M_{AB}$ then reads, in block form,
 \be M
= \left(\begin{array}{ccc}
\gamma^{-1} & \gamma^{-1}C & \gamma^{-1}\psi \\
C^T\gamma^{-1} & (\gamma+C^T)\gamma^{-1}(\gamma+C) &
(\gamma+C^T)\gamma^{-1}\psi\\
\psi^T\gamma^{-1} & \psi^T\gamma^{-1}(\gamma+C) & 1 +
\psi^T\gamma^{-1}\psi
\end{array}\right)\,,
\ee where $\gamma_{ab} \equiv e^{-\nu\phi}\lambda_{ab}$ ($\nu =
\sqrt{2/3}$), and $C$ is the $2\times2$ matrix $C = B +
\frac12\psi\psi^T$. The matrix $M$ is symmetric, and satisfies \be
MLM^T = L\,. \ee

The next step involves dualisation of the three--dimensional vector
fields according to \be\lb{dualsen}
\tau\sqrt{h}e^{\nu\phi}h^{ii'}h^{jj'}(ML)_{AB}F_{i'j'}^B =
\epsilon^{ijk}\partial_k\omega_A\,, \ee defines the row matrix \be
\omega \equiv (\ol\omega^a,\omega_a,\omega_5)\,. \ee Now, using the
result of \cite{Sen:1994wr} it is straightforward to write down the
$7\times 7$ matrix $\cM$ in a block form:
 \be\lb{Sen} {\cal M}
= \left(\begin{array}{ccc}
M+e^{-\nu\phi}\omega\omega^T & -e^{-\nu\phi}\omega^T&
ML\omega^T+\frac12 e^{-\nu\phi}\omega^T(\omega L\omega^T) \\
-e^{-\nu\phi}\omega & e^{-\nu\phi} &-\frac12 e^{-\nu\phi}(\omega
L\omega^T)\\
\omega LM+\frac12 e^{-\nu\phi}\omega (\omega L\omega^T)&-\frac12
e^{-\nu\phi}(\omega L\omega^T)&e^{\nu\phi}+\omega
LML\omega^T+\frac14 e^{-\nu\phi}(\omega L\omega^T)^2
\end{array}\right)\,.
\ee

Thus, in principle, the Sen's matrix can be also used in the case of
$D=6$ minimal supergravity not belonging to the sequence of the
heterotic string effective actions. But  disadvantage of such an
approach, apart from  relative complexity of the matrix (\ref{Sen}),
lies in the fact that  the variables of the five-dimensional
heterotic action in terms of which this representation is written,
are still non-trivially related to the initial six-dimensional
variables. Another desired feature which can be demanded from the
coset representation of the $D=6$ theory  is the possibility of its
simple truncation to five-dimensional minimal supergravity whose
$D=3$ coset $G_{2(2)}/(SU(2)\times SU(2))$ is an  invariant subspace
of the coset $O(4,3)/(O(4)\times O(3)) $. This can be achieved using
the direct toroidal compactification of $D=6$ minimal supergravity
to three dimensions. Before doing this, we briefly review the coset
structure of five-dimensional minimal supergravity. In both cases we
will use the technique applied in \cite{Bouchareb:2007ax} which
consists in i) obtaining an explicit form form of the target space
metric, ii) identifying its isometry algebra, iii) exponentiating
the Borel subalgebra to get suitable matrix representation. Though
technically different, it is conceptually the same construction as
used by Maharana-Schwarz and Sen \cite{Maharana:1992my,Sen:1994wr}.
\section{$D=5$ minimal supergravity}
The action of MSG5 reads \be S_{MSG5} = \int\left( \left[R -
\frac14F^{\mu\nu}F_{\mu\nu}\right]\sqrt{ g_{5} } -
\frac1{12\sqrt3}\epsilon^{\mu\nu\rho\sigma\lambda}F_{\mu\nu}
F_{\rho\sigma}A_{\lambda}\right)d^5x\,,
 \ee
with $F=dA$. We compactify on a two-torus using
 \ba
ds_{5}^2 &=& \lambda_{ab}(dz^a + a_i^adx^i)(dz^b
+ a_j^bdx^j) + \tau^{-1}h_{ij}dx^idx^j\,, \lb{ds53}\\
A_{(5)\mu} dx^\mu &=& \sqrt3(\psi_a dz^a + A_idx^i)\,,  \lb{A53} \ea
where $a,b=0,1$   and $\tau \equiv |{\rm det}\lambda|$. The $\nu =
i$ components of the Maxwell-Chern-Simons equations allow to dualize
the vector magnetic potential $A_i$ to a scalar magnetic potential
$\mu$ defined by
\begin{equation}\lb{dualmu}
F^{ij} = a^{aj} \partial^i \psi_a - a^{ai} \partial^j \psi_a +
\frac1{\tau \sqrt{h}} \epsilon^{ijk} \eta_k\,, \qquad \eta_k =
\partial_k \mu +   \epsilon^{ab} \psi_a \partial_k\psi_b \,.
\end{equation}
Similarly, the $\mu=i$, $\nu = a$ components of the Einstein
equations are integrated by
\begin{equation}\lb{dualom}
\lambda_{ab}G^{bij} =  \frac1{\tau \sqrt{h}} \epsilon^{ijk}
V_{ak}\,, \qquad V_{ak} =  \partial_k\omega_a -
\psi_a\left(3\partial_k\mu + \epsilon^{bc}
\psi_b\partial_k\psi_c\right)\,,
\end{equation}
where $G^b = da^b$, and $\omega_a$ is the `twist' or gravimagnetic
two-potential. The $D=3$ sigma model \be S_3=\int\left(
R_3(h)-\frac12 G_{AB}\partial_i\Phi^A
\partial_j\Phi^Bh^{ij}\sqrt{h}\right)d^3x\,,
 \ee
is then obtained with eight target space coordinates
$\Phi^A=\{\lambda_{ab},\omega_a, \psi_a, \mu\}$ and metric
 \be \lb{gtar}
dl^2=G_{AB}d\Phi^A d\Phi^B=\frac12 {\rm
Tr}(\lambda^{-1}d\lambda\lambda^{-1}d\lambda) +
\frac12\tau^{-2}d\tau^2 - \tau^{-1}V^T\lambda^{-1}V   +
3\left(d\psi^T\lambda^{-1}d\psi - \tau^{-1}\eta^2\right) \,.
 \ee
This space has 14 Killing vectors which were determined in terms of
these variables in \cite{Bouchareb:2007ax,5to3}. Nine manifest
infinitesimal symmetries (or generalised gauge transformations),
grouped according to their transformations under $GL(2R)$ (the group
of linear transformations in the $(z^1,z^2)$ plane) into the
quadruplet \be {M_a}^b =
2\lambda_{ac}\frac{\partial}{\partial\lambda_{cb}} +
\omega_a\frac{\partial}{\partial\omega_{b}} +
\delta_a^b\omega_c\frac{\partial}{\partial\omega_{c}} +
\psi_a\frac{\partial}{\partial\psi_{b}} +
\delta_a^b\mu\frac{\partial}{\partial\mu} \ee (the generators of the
$gl(2,R)$ subalgebra), the doublet and the singlet associated with
the the three cyclic `magnetic' coordinates: \be N^a  =
\frac{\partial}{\partial\omega_a}\,, \quad  Q =
\frac{\partial}{\partial\mu}\,, \ee and the doublet generating
infinitesimal gauge transformations of the $\psi_a$ \be R^a =
\frac{\partial}{\partial\psi_a} +
3\mu\frac{\partial}{\partial\omega_a} -
\epsilon^{ab}\psi_b\left(\frac{\partial}{\partial\mu} +
\psi_c\frac{\partial}{\partial\omega_c}\right)\,. \ee The five
remaining, non trivial infinitesimal isometries $L_a$, $P_a$ and $T$
closing the Lie algebra $g_{2(2)}$ are more complicated, their full
expression is given in \cite{5to3}. The $L_a$, ${M_a}^b$ and $N^a$
generate the vacuum subalgebra $sl(3,R)$. Assuming a spacelike
two-torus, the target space (\ref{gtar}) is identified as the coset
space $G_{2(2)}/(SU(2)\times SU(2))$.

The $7\times 7$ symmetric matrix representative of the coset
obtained by exponentiation of the Borel subalgebra
\cite{Bouchareb:2007ax,5to3} exhibits a highly nonlinear dependence
on the moduli. Its structure is quite different from that of the Sen
matrix (\ref{Sen}) for the coset $O(4,3)/(O(4)\times O(3))$, so it
is practically impossible to relate them.

\section{New representative for $D=6$ minimal supergravity}
A simpler representation of the coset $O(4,3)/(O(4)\times O(3)$ may
be achieved by performing direct compactification of the
six-dimensional theory on $T^3$. We start with the Lagrangian
(\ref{MSG6}), and assume 3 Killing vectors $\partial_a$ ($a =
1,2,3$). The six-dimensional metric and 3-form may be parameterized
by
 \ba\lb{6met}
ds\6^2 &=& \lambda_{ab}(dz^a + a_i^adx^i)(dz^b + a_j^bdx^j) +
\tau^{-1}h_{ij}dx^idx^j\,, \nn \\ \hat{G}_{abc} &=& 0\,, \quad
\hat{G}_{abi} = \hat{B}_{ab,i}\,,
 \ea
($\tau \equiv - \det  \lambda ,\; i,j = 4,5,6$) and the 10 remaining
components of $\hat{G}$ related to these by self-duality. Put
 \be\lb{bchi}
\hat{B}_{ab} \equiv \epsilon_{abc}\chi^c\,.
 \ee
Then,
 \be
\hat{G}_{abi} = \epsilon_{abc}\chi^c_{,i}\,, \quad \hat{G}^{aij} =
-\frac{\tau}{\sqrt{h}}\epsilon^{ijk}\chi^a_{,k}\,.
 \ee
The mixed Einstein equations
 \ba
\hat{R}^i_a &\equiv & \frac{\tau}{2\sqrt{h}}\partial_j[\tau\sqrt{h}
\lambda_{ab}{\cal F}^{bij}] \nn\\
&=& \frac12\hat{G}^{ibj}\hat{G}_{abj} =
\frac{\tau}{2\sqrt{h}}\partial_j[
\epsilon^{ijk}\epsilon_{abc}\chi^b_{,k}\chi^c]
 \ea
(${\cal F}^{b} \equiv da^b$) are solved by
 \be\lb{dual}
\lambda_{ab} {\cal F}^{bij} =
\frac1{\tau\sqrt{h}}\epsilon^{ijk}V_{ak}\,, \quad V_{ak} \equiv
\partial_k\omega_a + \epsilon_{abc}\chi^b_{,k}\chi^c\,.
 \ee
The remaining Einstein equations then lead to the gravitating sigma
model with target space metric
 \be\lb{tarmet}
dl^2 = \frac12 {\rm Tr}(\lambda^{-1}d\lambda\lambda^{-1}d\lambda) +
\frac12\tau^{-2}d\tau^2   - \tau^{-1}V^T\lambda^{-1}V -
2\tau^{-1}d\chi^T\lambda d\chi\,, \ee where \be V \equiv d\omega -
\chi\wedge d\chi\,.
 \ee

The dimension of this target space is twelve: six components of the
symmetric matrix $\lambda_{ab}$ and two triplets $\omega_a,\;
\chi^a$. In Appendix A we check that it admits 21 Killing vectors
generating the Lie algebra $o(4,3)$. These include nine Killing
vectors ${M_a}^b$ generating the algebra $gl(3,R)$ of linear
transformations in the three-Killing vector space, six vectors $N^a$
and $L_a$ which together with the ${M_a}^b$ generate the isometry
algebra $sl(4,R)$ for the target subspace corresponding to the
six-dimensional vacuum sector, and six more vectors $R_a$ and $P^a$
which complete the algebra $o(4,3)$. The fifteen Killing vectors
${M_a}^b$, $N^a$ and $R_a$ generate generalized gauge
transformations, with the $N^a$ generating translations of the
twists $\omega_a$ and the $R_a$ generating gauge transformations of
the $\chi^a$.

In Appendix B we construct real matrix representatives of $o(4,3)$,
beginning with the subalgebra $o(3,3) \sim sl(4,R)$. Rather than
using the Maison parametrization \cite{Maison:1979kx} of $sl(4,R)$
in terms of $4\times4$ matrices (which presumably would lead to a
representation of $o(4,3)$ in terms of $8\times8$ matrices), we use
the representation of $o(3,3)$ in terms of $6\times6$ matrices.
These are then promoted to $7\times7$ matrices by the addition of a
row and a column, and completed by six $7\times7$ matrices $R_a$ and
$P^a$ closing the algebra $o(4,3)$. The $7\times7$ coset matrix
representative is then constructed in a standard fashion as
 \be
{\cal M} = {\cal N}^T\eta{\cal N}\,,
 \ee
where ${\cal N}$ is obtained by exponentiating a suitable Borel
subalgebra of $0(4,3)$, and $\eta$ is a suitably chosen constant
matrix. The resulting coset representative is, in block form,
 \be\lb{M}
\M = \left(\begin{array}{ccc} \mu &
\sqrt{2}\mu\chi & \mu\gamma \\
\sqrt{2}\chi^T\mu & -1 + 2\chi^T\mu\chi&
\sqrt{2}(\chi^T\mu\gamma + \tilde{\chi}) \\
\gamma^T\mu & \sqrt{2}(\gamma^T\mu\chi + \tilde{\chi}^T) &
\gamma^T\mu\gamma - 2\tilde{\chi}^T\tilde{\chi} + \tilde{\mu}^{-1}
\end{array}\right)
 \ee
where $\tilde{}$ denotes the anti-transposition, i.e. transposition
relative to the anti- (or minor) diagonal, and
 \ba
&\mu  =  \tau^{-1}\lambda = \tau^{-1}\left(\begin{array}{ccc}
\lambda_{11} & \lambda_{12} & \lambda_{13} \\ \lambda_{21} &
\lambda_{22} & \lambda_{23} \\ \lambda_{31} & \lambda_{32} &
\lambda_{33} \end{array}\right) \,,& \nn\\
&\chi  =  \left(\begin{array}{c} \chi^1 \\ \chi^2
\\ \chi^3 \end{array}\right)\,, \quad
\tilde{\chi}=\left(\chi_3,\,\chi_2,\,\chi_1\right), \quad \gamma =
\hat{\omega} - \chi\tilde{\chi}\,, \quad \hat{\omega} =
\left(\begin{array}{ccc} -\omega_2 & \omega_3 & 0 \\ \omega_1 & 0 &
-\omega_3 \\ 0 & -\omega_1 & \omega_2
\end{array}\right)\,. &
 \ea
One can check that the target space metric (\ref{tarmet}) can be
expressed as
 \be
dl^2 = \frac14{\rm Tr}(\M^{-1}d\M\M^{-1}d\M)\,.
 \ee

In the case of a Lorentzian six-dimensional space $E_6$ with
signature $(-+++++)$ and an Euclidean reduced three-space (so that
one of the Killing vectors of $E_6$ is timelike), the symmetric
target space $\cal T$ of metric (\ref{tarmet}) is the coset $G/H =
O(4,3)/ O(2,2)\times O(2,1)$. $H$ is the isotropy group leaving
invariant any given point of the target space, which may be chosen
to be the point at infinity of $\cal T$. Thus it is relevant to
examine the various possible asymptotic behaviors for asymptotically
flat six-dimensional configurations.

{\em Minkowski asymptotics.} For an asymptotically Minkowskian
metric, or for a metric which is asymptotically the product of a
four-dimensional black hole by a 2-torus, with $x^1$ the time
coordinate, the asymptotic coset representative is \be \M_{\infty} =
\eta_M =
\left(\begin{array}{ccccccc} -1 & 0 & 0 & 0 & 0 & 0 & 0 \\
0 & 1 & 0 & 0 & 0 & 0 & 0 \\0 & 0 & 1 & 0 & 0 & 0 & 0 \\
0 & 0 & 0 & -1 & 0 & 0 & 0 \\0 & 0 & 0 & 0 & 1 & 0 & 0 \\
0 & 0 & 0 & 0 & 0 & 1 & 0 \\0 & 0 & 0 & 0 & 0 & 0 & -1
\end{array}\right)\,. \ee
This asymptotic behavior is preserved by the nine Killing vectors
 \ba \X_1 = -{M_2}^3 + {M_3}^2\,, & \quad \X_2 = {M_3}^1 + {M_1}^3\,,
& \quad \X_3 = -{M_1}^2 - {M_2}^1\,, \nn\\ \Y_1 = N^1 + L_1\,, &
\quad \Y_2 = N^2 - L_2\,, & \quad \Y_3 = N^3 - L_3\,, \nn\\ \Z_1 =
P^1 - R_1\,, & \quad \Z_2 = P^2 + R_2\,, & \quad \Z_3 = P^3 + R_3
 \ea
(with the first three pure gauge), satisfying the commutation
relations \ba && \left[\X_a,\X_b\right] = \left[\Y_a,\Y_b\right] =
\epsilon_{abc}\eta_c\X_c
\,, \quad \left[\Z_a,\Z_b\right] = 2\epsilon_{abc}\eta_c(\X_c+\Y_c)\,, \nn\\
&& \left[\X_a,\Y_b\right] =  \epsilon_{abc}\eta_c\Y_c\,, \quad
\left[\Y_a,\Z_b\right] = \left[\Z_a,\X_b\right] =
\epsilon_{abc}\eta_c\Z_c\,, \ea with $\eta_1 = -1, \eta_2 = \eta_3 =
+1$. The combinations \be K^0_a = \frac12(\X_a - \Y_a)\,, \quad
K^{\pm}_a = \frac14(\X_a + \Y_a \pm \Z_a) \ee generate three
commuting copies of the Lie algebra of $O(2,1)$,
 \be
\left[K_a,K_b\right] = \epsilon_{abc}\eta_cK_c\,.
 \ee
We thus recover the isotropy subgroup $H = O(2,2)\times O(2,1)$ =
$O(2,1)^3$.

{\em Black string asymptotics.} The static Myers-Perry (Tangherlini)
six-dimensional black string (the product of a five-dimensional
black hole by a circle) is
 \be ds_6^2 = -(1-m/r^2)dt^2 +
\frac{dr^2}{1-m/r^2} + \frac{r^2}4\left[(d\eta-\cos\theta
d\varphi)^2 + d\theta^2 + \sin^2\theta d\varphi^2\right] +
(d\zeta)^2\,.
 \ee
This has four commuting Killing vectors. Reduction relative e.g. to
$\partial_1 = \partial_t$, $\partial_2 =
\partial_{\eta}$ and $\partial_3 = \partial_{\zeta}$ leads to
 \ba
\lambda &=& \mbox{diag}[-(r^2-m)/r^2,\;r^2/4,\;1]\,, \quad \tau =
\frac{r^2-m}4\,, \quad a_{\varphi}^2 = -\cos\theta\,,  \nn \\
d\sigma^2 &\equiv& h_{ij}dx^idx^j =
\frac{r^2}4\left[dr^2+\frac{r^2-m}4(d\theta^2 + \sin^2\theta
d\varphi^2)\right]\,,
 \ea
leading to \be \omega_2 = \frac{r^2+c}4\,, \ee with $c$ a constant
of integration. Computation of the asymptotic behavior of the lower
right-hand side $3\times3$ block in (\ref{M}) gives \be
\tau^{-1}\gamma^T\lambda\gamma + \tau\tilde{\lambda}^{-1}
\stackrel{\simeq}{_{r\to\infty}}
\mbox{diag}[-(m+2c)/4,\;1,\;(m+2c)/4]\,, \ee which is equal to the
asymptotic behavior of the upper left-hand side block for the value
$c = -m/2$. In this case, \be \M_{\infty} =
\left(\begin{array}{ccccccc} 0 & 0 & 0 & 0 & 1 & 0 & 0 \\
0 & 1 & 0 & 0 & 0 & 0 & 0 \\0 & 0 & 0 & 0 & 0 & 0 & 1 \\
0 & 0 & 0 & -1 & 0 & 0 & 0 \\1 & 0 & 0 & 0 & 0 & 0 & 0 \\
0 & 0 & 0 & 0 & 0 & 1 & 0 \\0 & 0 & 1 & 0 & 0 & 0 & 0
\end{array}\right)\,. \ee
This asymptotic behavior is preserved by the nine Killing vectors
\ba\lb{hstring} X_+ = - {M_2}^1 - L_1 \,,& \quad X_0 = {M_2}^2\,,&
\quad
X_- = - {M_1}^2 + N^1\,, \nn\\
Y_+ = {M_2}^3 - L_3 \,,& \quad Y_0 = {M_1}^3 + {M_3}^1\,,& \quad
Y_- = {M_3}^2 + N^3\,, \nn\\
Z_+ = -P^1 + P^3 \,, & \quad Z_0 = P^2 + R_2\,, &\quad Z_- = -R_1 +
R_3\,. \ea $X_0$, $X_-$, $Y_0$, $Y_-$ and $Z_-$ are pure gauge. The
first three generate an $SL(2,R)$, \be \left[X_0,X_{\pm}\right] =
\pm X_{\pm}\,, \quad \left[X_+,X_-\right] = X_0\,, \ee or
symbolically $\left[X,X\right] = X$. The full algebra \ba &&
\left[X,X\right] = \left[Y,Y\right] = X
\,, \quad \left[Z,Z\right] = 2(X+Y)\,, \nn\\
&& \left[X,Y\right] =  \left[Y,X\right] = Y\,, \quad
\left[X,Z\right] = \left[Z,X\right] = \left[Y,Z\right] =
\left[Z,Y\right] = Z\,, \ea (with commutators such as
$\left[X_0,Y_0\right]$ and $\left[X_{\pm},Y_{\pm} \right]$
vanishing) can be split, as in the case of Minkowski asymptotics,
into three commuting $sl(2,R) = O(2,1)$ generated by the
combinations \be J^0 = \frac12(X - Y)\,, \quad J^{\pm} = \frac14(X +
Y \pm Z)\,, \ee so that the isotropy subgroup $H = O(2,2)\times
O(2,1)$ is again recovered.

{\em Black hole asymptotics.} The static six-dimensional black hole
 \be ds_6^2 = -(1-m/r^3)dt^2 + \frac{dr^2}{1-m/r^3}
 + r^2\left[d\theta^2 + \cos^2\theta d\zeta^2 + \sin^2\theta
(d\eta^2 + \sin^2\eta d\varphi^2)\right]
 \ee
has only three commuting Killing vectors $\partial_t$,
$\partial_{\zeta}$ and $\partial_{\varphi}$. Reduction relative to
these vectors leads to
 \ba
\mu &=& \mbox{diag}[-4/r^4\sin^22\theta\sin^2\eta,\;
r/(r^3-m)\sin^2\theta\sin^2\eta,\;r/(r^3-m)\cos^2\theta]\,. \nn \\
\chi &=& \gamma = 0\,.
 \ea
The resulting matrix $\cal M$ has no regular limit at spatial
infinity. As in the well-known case of four-dimensional
Einstein-Maxwell theory reduced relative to the azimuthal Killing
vector \cite{ErnstWild76}, new solutions can be generated from this
by $O(4,3)$ transformations, but these will be always
non-asymptotically flat.

\section{Invariant subspaces}
The new matrix (\ref{M}) parameterizes  the twelve-dimensional coset
space of MSG6 theory. It may be also applied as a representative of
the embedded eight-dimensional coset $G_{2(2)}/(SU(2)\times SU(2)$
corresponding to MSG5 and four-dimensional coset
$SU(2,1)/S(U(2)\times U(1))$ corresponding to EM4. These may be
selected by purely algebraic constraints on the potentials. To find
these constraints one has to consider dimensional reductions and
consistent truncations which relate these theories to MSG6.
\subsection{$D=5$ minimal supergravity}
The coset $ G_{2(2)}/(SU(2)\times SU(2)$ is a totally geodesic
subspace of the coset $O(4,3)/(O(4)\times O(3))$, so MSG6
compactified on a circle can be consistently truncated to MSG5.
Indeed, it can be checked \cite{oxi} that the equations of motion
following from (\ref{O435}) are consistent with the constraints
 \be\lb{O43G_2}
\phi=0\,,\qquad
F_{\mu\nu}=\mp\frac{\sqrt{-g_5}}{3\sqrt{2}}\epsilon_{\mu\nu\rho\sigma\tau}
H^{\rho\sigma\tau}\,,
 \ee
in which case they reduce to those of MDG5, provided the two-form
field $F_{\mu\nu}$ is rescaled by
 \be\lb{rescal}
F_{\mu\nu} \to \pm\sqrt{\frac23}F_{\mu\nu}\,.
 \ee
Note that, in view of the second relation in (\ref{56rel}), the
second contraint (\ref{O43G_2}) is equivalent to identification of
the Maxwell two-form and the Kaluza-Klein two-form of the reduced
theory (\ref{O435}), in which case the dilaton can be consistently
set to zero. We must now identify the corresponding constraints in
terms of the target space variables. Inspecting the definitions of
the three-dimensional target space variables
(\ref{6met})-(\ref{dual}) one finds
 \be\lb{cons3}
\lambda_{33} = 1\,, \quad \lambda_{a3} = \mp\epsilon_{ab}\chi^b\,,
\quad \omega_3 = \mp\chi^3\,.
 \ee
Since the $G_2$ sector arises as a consistent truncation of the
five-dimensional reduction of the original six-dimensional model, it
is not surprising that in the reduction to three dimensions the
three-covariance is broken down to two-covariance.  Actually,
knowing this (and making some educated guesses) is enough to find
the two $g_2$ subalgebras generated by the Killing vectors
preserving the constraints (\ref{cons3}). In the two-covariant
notation of Sect. 4 of \cite{Bouchareb:2007ax}, their generators are
related to those of $o(4,3)$ by
 \be\lb{o43g2}
\begin{array}{ccl}
\underline{g_2} & & \underline{o(4,3)} \vspace{2mm} \\
{M_a}^b &= & \;\;\;{M_a}^b \\
N^a &= & \;\;\;N^a \\
L_a &=& \;\;\;L_a \\
Q &= & - N^3 \pm R_3 \\
T &= & - L_3 \mp P^3\,, \\
R^a &= & - {M_3}^a \pm \epsilon^{ab}R_b \\
P_a &= & - {M_a}^3 \pm \epsilon_{ab}P^b
\end{array}
 \ee
with $a,b = 1,2$. It is easy to check that these combinations
satisfy the commutation relations (92)-(97) of
\cite{Bouchareb:2007ax}.

Conversely, comparing the covariant three-dimensional reductions of
MSG5 and MSG6, we see that any solution ($ds_{(5)}^2$, $A_{(5)}$) of
MSG5 with two commuting Killing vectors can be oxidized to a
solution of MSG6 with three commuting Killing vectors given by
 \ba\lb{MSG56}
ds_{(6})^2 &=& ds_{(5)}^2 + (- \psi_a dx^a + dz - A_i dx^i)^2\,, \nn\\
\chi &=& \pm(-\psi_2,\;\psi_1,\;\mu)\,.
 \ea
It follows that, given a solution of MSG5 with two commuting Killing
vectors, one can generate from this a new solution by going through
the following steps: 1) oxidize the seed solution to a solution of
MSG6 by (\ref{MSG56}); 2) construct its coset representative
(\ref{M}); 3) transform this,
 \be
{\cal M'} = P^T{\cal M}P\,,
 \ee
by the action of an $O(4,3)$ transformation $P$ generated by the
generators of the second column of (\ref{o43g2}); 4) extract from
${\cal M'}$ the new solution of MSG6; 5) reduce this to five
dimensions by (\ref{MSG56}). In view of the simplicity of the matrix
representation (\ref{M}) compared to that previously known for MSG5,
this procedure might be easier to implement than direct generation
by $G_{2(2)}$ transformations.

The generators preserving both five-dimensional Myers-Perry (or
black string) asymptotics {\em and} $G_2$ truncation are
 \be
J^{\mp} = \frac14(X + Y \mp Z)\,, \quad J^0 + J^{\pm} = \frac14(3X -
Y \pm Z)\,,
 \ee
generating two commuting copies of $sl(2,R) = o(2,1)$. The
non-trivial generators are
 \ba \pm G_{0(\pm)} &=& Z_0 \mp Y_0 = P^2
+ R_2 \mp
({M_1}^3 + {M_3}^1)\qquad (\mbox{\rm electric charge})\,, \nn\\
\pm G_{+(\pm)} &=& Z_+ \mp Y_+ = -P^1 + P^3 \mp({M_2}^3 - L_3)
\qquad
(\mbox{\rm two dipole charges})\,, \nn\\
F_+ &=& X_+ = - {M_2}^1 - L_1 \qquad (\mbox{\rm angular
momentum})\,.
 \ea
Clearly, all the generators (\ref{hstring}) preserving black string
asymptotics are linear combinations of the four non-trivial
generators $F_+$, $G_{+(+)}$, $G_{+(-)}$, and one of the
$G_{0(\pm)}$, together with gauge transformations. These four
generators applied to a black string will do the same job as the
corresponding $G_2$ generators (and in particular, preserve the
black string condition $\lambda_{33}=1§$), but in a simpler fashion.
The non-vanishing commutators between these four generators are \ba
\left[G_{0(\pm)},F_+\right] &=& G_{+(\pm)} \,, \nn \\
\left[G_{0(\pm)},G_{+(\pm)}\right] &=& 3F_+ - 2G_{+(\pm)} \,, \\
\left[G_{0(\pm)},G_{+(\mp)}\right] &=& F_+ - G_{+(+)} - G_{+(-)} \,.
\nn \ea
\subsection{$D=4$ Einstein-Maxwell }
To identify the constraints selecting the $SU(2,1)/S(U(2)\times
U(1))$ subspace of the $G_2$ coset, one must first compactify MSG5
on a circle\cite{Bouchareb:2007ax}, since this subcoset corresponds
to four-dimensional Einstein-Maxwell theory.  Assuming the existence
of a space-like Killing vector $\partial_z$, we parametrize the
five-dimensional metric and Maxwell field by
\begin{eqnarray}
ds\5^2 &=& \mathrm{e}^{-2\phi } (dz + C_\mu dx^\mu)^2 +
\mathrm{e}^{\phi} ds\4^2,\label{ds5}\\
A\5 &=& A_\mu dx^\mu + \sqrt{3}\kappa dz,
\end{eqnarray}
($\mu = 1\ldots 4$). The corresponding four-dimensional action
\begin{equation}
S_4 =   \int d^4x \sqrt{-g} \left[ R - \frac32 (\partial \phi)^2 -
\frac32 \mathrm{e}^{2\phi} (\partial \kappa)^2 - \frac14
\mathrm{e}^{-3\phi} G^2 - \frac14\mathrm{e}^{-\phi} \tilde F^2 -
\frac12 \kappa F F^* \right],
\end{equation}
where
\begin{equation}
  G = dC, \quad F = dA, \quad \tilde F = F +
\sqrt{3} C \wedge d\kappa,
\end{equation} and $F^*$ is the four-dimensional Hodge dual of $F$, describes
an Einstein theory with two coupled abelian gauge fields $F$ and
$G$, a dilaton $\phi$ and an axion $\kappa$. The field equations in
terms of the four-dimensional variables read
\begin{eqnarray}
\nabla^2 \phi - \mathrm{e}^{2\phi} (\partial \kappa)^2 + \frac14
\mathrm{e}^{-3\phi} G^2 + \frac1{12} \mathrm{e}^{-\phi} \tilde F^2
&=& 0,
\\
\nabla_\mu \left( \mathrm{e}^{2\phi} \nabla^\mu \kappa \right) -
\frac1{3} \left[ \sqrt3\,  \nabla_\mu (\mathrm{e}^{-\phi} \tilde
F^{\mu\nu} C_\nu) + \frac12 F_{\mu\nu} F^{*\mu\nu} \right] &=& 0,
\\
\nabla_\mu \left( \mathrm{e}^{-\phi} \tilde F^{\mu\nu} + 2\kappa
F^{*\mu\nu} \right) &=& 0,
\\
\nabla_\mu \left( \mathrm{e}^{-3\phi} G^{\mu\nu} \right) +  \sqrt3\,
\mathrm{e}^{-\phi} \tilde F^{\mu\nu} \partial_\mu \kappa &=& 0\,.
\end{eqnarray}
Truncation to the Einstein-Maxwell system is achieved by imposing
 \be
\phi = 0\,, \quad \kappa = 0\,, \quad G_{\mu\nu} = \frac1{2\sqrt3}
\sqrt{ g_4 }\epsilon_{\mu\nu\rho\sigma}F_{(4)}^{\rho\sigma}\,.
 \ee
After reduction to three dimensions, this leads to the constraints
 \be
\lambda_{22} = 1\,, \quad \psi_2 = 0\,, \quad \lambda_{12} = \mu\,,
\quad \omega_2 = - \psi_1\,.
 \ee
We find that these constraints are preserved by the eight
infinitesimal transformations \ba
& \qquad K_1 = {M_1}^1 \,, & \nn \\
& \begin{array}{ll}K_2 = {M_2}^1 + Q \,, & K_3 = {M_1}^2 - T \,,\nn \\
K_4 = N^1 \,, \;& K_5 = L_1 \,,\nn \\
K_6 = N^2 - R^1 \,, \;& K_7 = L_2 + P_1 \,,\end{array}\nn \\
&K_8 = P_2 - R^2\,. & \ea From the commutation relations given in
\cite{5to3}, we find that the $K_M$ ($M = 1,...,8$) generate the Lie
algebra of $SU(2,1)$, which may be put in the Cartan-Weyl form
\cite{gilmore}, with \ba
&H_1 = \frac1{2\sqrt3}K_1\,, \quad H_2 = \frac{i}6K_8\,,& \nn\\
&\begin{array}{lll}E_1 = \frac1{\sqrt6}K_5\,, & E_{-} =
-\frac1{\sqrt6}K_4\,,
& \alpha_1 = \frac1{\sqrt3}(1,0)\,,\\
E_2 = \frac1{4\sqrt3}(-K_6+iK_2)\,, & E_{-2} =
\frac1{4\sqrt3}(K_7-iK_3)
\,,& \alpha_2 = \frac1{\sqrt3}\bigg(-\frac12,\frac{\sqrt3}2\bigg)\,,\\
E_3 = \frac1{4\sqrt3}(K_3-iK_7)\,, & E_{-3} =
\frac1{4\sqrt3}(K_2-iK_6)\,,& \alpha_3 =
\frac1{\sqrt3}\bigg(\frac12,\frac{\sqrt3}2\bigg)\,.\end{array}& \ea

\section{Outlook}
The main result of this paper is the new representative (\ref{M}) of
the coset $O(4,3)/(O(4)\times O(3))$ which is a symmetric $7\times
7$ matrix, given in block form. This matrix is substantially simpler
than the matrix (\ref{Sen}) constructed by Sen's method. Moreover,
it also looks simpler than the $G_2$ matrix for the coset
$G_{2(2)}/(SU(2)\times SU(2))$ constructed by us previously
\cite{Bouchareb:2007ax} and used for solution generation in
\cite{Bouchareb:2007ax,TYM}. The reason is that constraining to the
subspace $G_{2(2)}/(SU(2)\times SU(2))$ one loses the $O(4,3)$
covariance which simplifies the underlying matrix structure. After
truncation to vacuum five-dimensional gravity ($\chi^a=0$), the new
matrix leads to a $7\times 7$ matrix representative of the coset
$SL(3,R)/O(2,1)$ which is different from that resulting from the
truncation to vacuum gravity ($\psi_a=\mu=0$) of our previous $G_2$
matrix \cite{Bouchareb:2007ax}. It is therefore expected that
imposing the constraints (\ref{cons3}) on the coordinates of the
full coset $O(4,3)/(O(4)\times O(3))$ one should obtain a $7\times
7$ representative of the $G2$ coset different from our previous one.
This new $G2$ matrix could also be obtained as in
\cite{Bouchareb:2007ax,5to3} by direct exponentiation of the Borel
subalgebra using the new representation (\ref{o43g2}),
(\ref{MNL})-(\ref{matRP}) of the $g2$ algebra. Alternatively, one
can as we have shown use the full new matrix (\ref{M}) together with
the corresponding transformations (\ref{o43g2}) to generate from a
given seed a new solution of five-dimensional minimal supergravity.

At the same time, one can also transform solutions of MSG5 to
non-trivial solutions of MSG6 by performing $O(4,3)$ transformations
which do not belong to the $G_{2(2)}$ subgroup, which may be chosen
to have required asymptotic properties, as discussed at the end of
section 4. The same arguments equally apply to the
$SU(2,1)/S(U(2)\times U(1))$ subspace of the $G2$ coset.

\acknowledgments The authors are grateful to Adel Bouchareb,
Chiang-Mei Chen, C\'edric Leygnac and Nicolai Scherbluck for
fruitful collaboration. The work of subsection 5B was carried out in
collaboration with Adel Bouchareb. D.G. is grateful to the LAPTH
Annecy for hospitality in December 2012 while the paper was written.
His work was supported in part by the RFBR grant 11-02-01371-a.

\appendix
\section{Isometry algebra of the metric (\ref{tarmet})}

Fifteen obvious Killing vectors are: \be {M_a}^b =
2\lambda_{ac}\frac{\partial}{\partial\lambda_{cb}} +
\omega_a\frac{\partial}{\partial\omega_{b}} +
\delta_a^b\omega_c\frac{\partial}{\partial\omega_{c}} -
\chi^b\frac\partial{\partial\chi^a} + \delta^b_a
\chi^c\frac\partial{\partial\chi^c}\,, \ee generating linear
transformations in the three-Killing vector space, \be N^a =
\frac{\partial}{\partial\omega_a}\,, \ee generating translations of
the ``magnetic'' coordinates $\omega_a$, and \be R_a =
\frac{\partial}{\partial\chi^a} +
\epsilon_{abc}\chi^b\frac{\partial}{\partial\omega_c}\,, \ee
generating gauge transformations of the $\chi^a$.

Their commutation relations are \ba \left[{M_a}^b,{M_c}^d\right] &=&
\delta_c^b{M_a}^d - \delta_a^d{M_c}^b\,,
\lb{MM}\\
\left[{M_a}^b,N^c\right] &=& -\delta_a^cN^b - \delta_a^bN^c\,, \lb{MN}\\
\left[{M_a}^b,R_c\right] &=& \delta_c^bR_a - \delta_a^bR_c \,, \lb{MR}\\
\left[N^a,N^b\right] &=& 0\,,   \lb{NN}\\
\left[N^a,R_b\right] &=& 0\,, \lb{NR}\\
\left[R_a,R_b\right] &=& -2\epsilon_{abc}N^c\,. \lb{RR} \ea

Three more vectors $L_a$ are needed to complete the algebra
$sl(4,R)$ of the vacuum sector: \ba
\left[{M_a}^b,L_c\right] &=& \delta_c^bL_a + \delta_a^bL_c\,,\lb{ML}\\
\left[N^a,L_b\right] &=& {M_b}^a\,, \lb{NL}\\
\left[L_a,L_b\right] &=& 0\,. \lb{LL} \ea Adding to the known form
of the $sl(4,R)$ for 6D Einstein the information from (\ref{NL}),
\ba L_a &=& \omega_a\omega_b\frac{\partial}{\partial\omega_b} +
2\omega_b\lambda_{ac}\frac{\partial}{\partial\lambda_{bc}} +
\chi^b(\omega_a\frac{\partial}{\partial\chi^b} -
\omega_b\frac{\partial}{\partial\chi^a}) + \tau\lambda_{ab}
\frac{\partial}{\partial\omega_b} + \cdots \ea (the omitted terms
are of order 0 in $\omega_a$). Assuming that the full Lie algebra is
$O(4,3)$, it must close with the three remaining generators $P^a$
defined by \be\lb{RL} \left[R_a,L_b\right] = \epsilon_{abc}P^c \,,
\ee leading to \be P^a =
\omega_b(\chi^b\frac{\partial}{\partial\omega_a} -
\chi^a\frac{\partial}{\partial\omega_b} - \epsilon^{abc}
\frac{\partial}{\partial\chi_c})  + \cdots\,, \ee and obeying the
commutation relations \ba
\left[{M_a}^b,P^c\right] &=& -\delta^c_aP^b + \delta^b_aP^c \,, \lb{MP}\\
\left[{N^a},P^b\right] &=& \epsilon^{abc}R_c \,, \lb{NP}\\
\left[{R_a},P^b\right] &=& 2{M_a}^b - \delta_a^b{\rm Tr}(M) \,,\lb{RP}\\
\left[{L_a},P^b\right] &=& 0\,, \lb{LP}\\
\left[{P^a},P^b\right] &=& -2\epsilon^{abc}L_c\,. \lb{PP} \ea

The degrees of the various fields can be found from their
commutators with ${\rm Tr}(M)$; \be [\lambda] = 2\,, \quad [\omega]
= 4\,, \quad [\chi] = 2\,. \ee This leads to the degrees of the
various Killing vectors \be
 \left[{M_a}^b\right] = 0\,,\quad
\left[R_a\right] = -2\,,\quad  \left[P^b\right] = 2\,, \quad
\left[N^b\right] = -4\,,\quad  \left[L_a\right] = 4\,. \ee

The six unknown Killing vectors $L_a$ and $P^a$ can be determined,
up to a sign, by solving the commutation relations (\ref{RP}) and
(\ref{PP}). The relatively simple result is \ba L_a &=&
\omega_a\omega_b\frac{\partial}{\partial\omega_b} +
2\omega_b\lambda_{ac}\frac{\partial}{\partial\lambda_{bc}} +
\chi^b\left(\omega_a\frac{\partial}{\partial\chi^b} -
\omega_b\frac{\partial}{\partial\chi^a}\right) + \tau\lambda_{ab}
\frac{\partial}{\partial\omega_b} \nn\\ &&
-2\epsilon_{abc}\chi^b\chi^d\lambda_{de}\frac{\partial}{\partial\lambda_{ec}}
-\alpha\tau\epsilon_{abc}\lambda^{bd}\chi^c
\left(\frac{\partial}{\partial\chi^d} -
\epsilon_{def}\chi^e\frac{\partial}{\partial\omega_f}\right) \,,\\
P^a &=& \omega_b\left(\chi^b\frac{\partial}{\partial\omega_a} -
\chi^a\frac{\partial}{\partial\omega_b} - \epsilon^{abc}
\frac{\partial}{\partial\chi_c}\right) +
2\chi^b\left(2\lambda_{bc}\frac{\partial}{\partial\lambda_{ca}} -
\delta^a_b\lambda_{dc}\frac{\partial}{\partial\lambda_{cd}}\right)
\nn\\ && -\chi^a\chi^b\frac{\partial}{\partial\chi_b} -
\alpha\tau\lambda^{ab}\left(\frac{\partial}{\partial\chi^b} -
\epsilon_{bcd}\chi^c\frac{\partial}{\partial\omega_d}\right) \,, \ea
with $\alpha^2 = 1$.

The value of $\alpha = \pm1$ is presumably related to the signature
of $\lambda$ (here $-++$). It can be determined by enforcing that
e.g. $\gamma_aP^a$ ($\gamma_a$ constant vector) is a Killing vector
of the target space metric. The action of $(P\gamma)$ leads to the
first order variations (written in matrix notation) \ba
\delta\lambda &=& 2\left[\gamma\cdot\chi\lambda +
\lambda\chi\cdot\gamma -
(\chi\gamma)\lambda\right]\,, \nn\\
\delta\omega &=& \gamma(\chi\omega) - (\chi\gamma)\omega
+ \alpha\tau\lambda^{-1}\gamma\wedge\chi\,, \\
\delta\chi &=& -\gamma\wedge\omega - (\chi\gamma)\chi -
\alpha\tau\lambda^{-1} \gamma\,. \nn \ea This leads to \ba
\delta(dl^2) &=& 4(1-\alpha)\left[(d\chi d\lambda\lambda^{-1}\gamma)
- \tau^{-1}(d\tau+(\chi\lambda d\chi))(d\chi\gamma)\right. \nn\\ &&
\left. +\tau^{-1}(\chi\gamma)(d\chi\lambda d\chi)
-\tau^{-1}(\gamma,\lambda d\chi,d\omega)\right]\,, \ea which
vanishes provided \be \alpha = +1\,. \ee

\renewcommand{\theequation}{B.\arabic{equation}}
\section{Matrix representative}

The first step is to construct real matrix representatives of
$O(4,3)$, beginning with the subalgebra $O(3,3) \sim sl(4,R)$.
Rather than using the Maison parametrisation of $sl(4,R)$ in terms
of $4\times4$ matrices (which presumably would lead to a
representation of $O(4,3)$ in terms of $8\times8$ matrices), we use
the representation of $O(3,3)$ in terms of $6\times6$ matrices,
decomposed in $3\times3$ blocks according to \be {M_a}^b =
\left(\begin{array}{cc} {m_a}^b & 0 \\ 0 & -\tilde{m}_a^{\,b}
\end{array}\right)\,, \quad N^a =
\left(\begin{array}{cc} 0 & n^a \\ 0 & 0\end{array}\right)\,, \quad
L_a = \left(\begin{array}{cc} 0 & 0 \\ - n^{aT} & 0
\end{array}\right)\,, \ee where $\tilde{}$ denotes the
anti-transposition, i.e. transposition relative to the anti- (or
minor) diagonal, and \ba && {({m_a}^b)^{\alpha}}_{\beta} =
\delta_a^{\alpha}\delta^b_{\beta} -
\delta_a^b\delta^{\alpha}_{\beta}\,,
\\ && n^1 = \left(\begin{array}{ccc} 0 & 0 & 0 \\ 1 & 0 & 0 \\ 0 &
-1 & 0
\end{array}\right)\,, \quad n^2 = \left(\begin{array}{ccc} -1 & 0 &
0 \\ 0 & 0 & 0 \\ 0 & 0 & 1 \end{array}\right)\,, \quad n^3 =
\left(\begin{array}{ccc} 0 & 1 & 0 \\ 0 & 0 & -1 \\ 0 & 0 & 0
\end{array}\right) \nn
\ea ($\alpha,\beta=1,2,3$). These matrices satisfy the commutation
relations (\ref{MM}), (\ref{MN}), (\ref{ML}), (\ref{NN}), (\ref{NL})
and (\ref{LL}).

The $7\times7$ matrix generators of $O(4,3)$ contain the preceding,
promoted to $7\times7$ matrices by the addition of a central 3-row
and a central 3-column, in block form
 \be\lb{MNL} {M_a}^b =
\left(\begin{array}{ccc} {m_a}^b & 0 & 0 \\ 0 & 0 & 0 \\ 0 & 0 &
-\tilde{m}_a^{\,b}
\end{array}\right)\,, \quad N^a =
\left(\begin{array}{ccc} 0 & 0 & n^a \\ 0 & 0 & 0 \\ 0 & 0 & 0
\end{array}\right)\,, \quad L_a = \left(\begin{array}{ccc} 0 & 0 & 0
\\ 0 & 0 & 0 \\ - n^{aT} & 0 & 0
\end{array}\right)\,,
 \ee
together with
 \be\lb{matRP}
R_a = \sqrt2\left(\begin{array}{ccc} 0 & r_a & 0 \\
0 & 0 & -\tilde{r}_a \\ 0 & 0 & 0
\end{array}\right)\,, \quad P^a = \sqrt2\left(\begin{array}{ccc} 0 & 0 & 0
\\ r_a^T & 0 & 0 \\ 0 & - \tilde{r}_a^T & 0 \end{array}\right)\,,
 \ee
where $r_a$ is the column matrix of elements \be r_a^{\alpha} =
\delta_a^{\alpha}\,. \ee Using \be r_a\tilde{r}_b - r_b\tilde{r}_a =
\epsilon_{abc}n^c\,, \ee these can be checked to satisfy the
remaining commutation relations of $O(4,3)$.

The $7\times7$ coset matrix representative is \be {\cal M} = {\cal
V}^T\M_0{\cal V}\,, \ee with \be\lb{M0} {\cal M}_0 =
\left(\begin{array}{ccc} \mu  & 0 & 0 \\ 0 & -1 & 0 \\ 0 & 0 &
\tilde{\mu}^{-1} \end{array}\right)\,, \quad \mu =
\tau^{-1}\lambda\,, \ee and \be {\cal V} =
e^{\chi^aR_a}e^{\omega_aN^a} = \left(\begin{array}{ccc} 1 &
\sqrt2\chi & \gamma
\\ 0 & 1 & -\sqrt2\tilde{\chi} \\ 0 & 0 & 1
\end{array}\right)\,, \ee
where
 \be\lb{chigaom} \chi = \left(\begin{array}{c} \chi^1 \\ \chi^2
\\ \chi^3 \end{array}\right)\,, \quad \gamma = \hat{\omega} - \chi\tilde{\chi}\,,
\quad \hat{\omega} = \left(\begin{array}{ccc} -\omega_2 & \omega_3 &
0 \\ \omega_1 & 0 & -\omega_3 \\ 0 & -\omega_1 & \omega_2
\end{array}\right)\,.
 \ee
The resulting coset representative \be \M = \left(\begin{array}{ccc}
\mu &
\sqrt{2}\mu\chi & \mu\gamma \\
\sqrt{2}\chi^T\mu & -1 + 2\chi^T\mu\chi&
\sqrt{2}(\chi^T\mu\gamma + \tilde{\chi}) \\
\gamma^T\mu & \sqrt{2}(\gamma^T\mu\chi + \tilde{\chi}^T) &
\gamma^T\mu\gamma - 2\tilde{\chi}^T\tilde{\chi} + \tilde{\mu}^{-1}
\end{array}\right) \ee is related to its inverse by \be \M^{-1} = \tilde{\M} \ee (use
$\tilde{\cal V}(\omega,\chi) = {\cal V}(-\omega,-\chi)$). Taking
into account the identity $${\rm Tr}[\tilde{\lambda}V^T\lambda V] =
-2\tau(V^T\lambda^{-1}V)\,,$$ which follows from (\ref{J31}), one
checks that the target space metric (\ref{tarmet}) can be expressed
as \be dl^2 = \frac14{\rm Tr}(\M^{-1}d\M\M^{-1}d\M)\,. \ee

The Kaluza-Klein vectors $a_i^a$ can be recovered directly by
solving the duality equation (\ref{dual}), where the field $V$ is
contained in the block
 \be\lb{J31}
{\cal J}_{31} = \tau^{-2} \tilde{\lambda}(d\hat{\omega} + d\chi
\tilde{\chi} - \chi d\tilde{\chi})^T\lambda = -
\tau^{-1}(\widehat{\lambda^{-1}V})^T
 \ee
of the current
 \be {\cal J} = \M^{-1}d\M
 \ee
(with the hat $\hat{}$ vector-to-matrix transformation defined as in
the last equation (\ref{chigaom})).
%\newpage


\begin{thebibliography}{99}

\bibitem{GG} M. Gunaydin and F. G\"{u}rsey: J. Math. Phys. \textbf{14},
1651 (1973).

 %\cite{Gunaydin:2007qq}
\bibitem{Gunaydin:2007qq}
  M.~Gunaydin, A.~Neitzke, O.~Pavlyk and B.~Pioline,
  %``Quasi-conformal actions, quaternionic discrete series and twistors: SU(2,1) and G_2(2),''
  Commun.\ Math.\ Phys.\  {\bf 283}, 169 (2008)
  [arXiv:0707.1669 [hep-th]].
  %%CITATION = ARXIV:0707.1669;%%

%\cite{Cremmer:1999du}
\bibitem{Cremmer:1999du}
  E.~Cremmer, B.~Julia, H.~Lu and C.~N.~Pope,
  ``Higher dimensional origin of D = 3 coset symmetries'',
  hep-th/9909099.
  %%CITATION = HEP-TH/9909099;%%

\bibitem{mizoh} S.~Mizoguchi and N.~Ohta, Phys.\ Lett.\  B {\bf
441} (1998) 123 [hep-th/9807111].

%\cite{Bouchareb:2007ax}
\bibitem{Bouchareb:2007ax}
  A.~Bouchareb, G.~Cl\'ement, C.~-M.~Chen, D.~V.~Gal'tsov, N.~G.~Scherbluk, T.~Wolf,
  %``G(2) generating technique for minimal D=5 supergravity and black rings,''
  Phys.\ Rev.\  {\bf D76}, 104032 (2007).
  [arXiv:0708.2361 [hep-th]].

\bibitem{Ernst:1967wx}
  F.~J.~Ernst,
  %``New formulation of the axially symmetric gravitational field problem,''
  Phys.\ Rev.\  {\bf 167}, 1175 (1968);
  %%CITATION = PHRVA,167,1175;%%
  %\cite{Ernst:1967by}
%\bibitem{Ernst:1967by}
  F.~J.~Ernst,
  %``New Formulation of the Axially Symmetric Gravitational Field Problem. II,''
  Phys.\ Rev.\  {\bf 168}, 1415 (1968).
  %%CITATION = PHRVA,168,1415;%%

%\cite{Kinnersley:1977pg}
\bibitem{Kinnersley:1977pg}
  W.~Kinnersley,
  %Generation of Stationary Einstein-Maxwell Fields.
  J. Math. Phys. 14: 651-653 (1973);
  %``Symmetries of the Stationary Einstein-Maxwell Field Equations. 1.,''
  J.\ Math.\ Phys.\  {\bf 18}, 1529 (1977).
  %%CITATION = JMAPA,18,1529;%%
  %\cite{Kinnersley:1977ph}
%\bibitem{Kinnersley:1977ph}
 % W.~Kinnersley and D.~M.~Chitre,
  %``Symmetries of the Stationary Einstein-Maxwell Field Equations. 2.,''
  %J.\ Math.\ Phys.\  {\bf 18}, 1538 (1977).
  %%CITATION = JMAPA,18,1538;%%
  %\cite{Kinnersley:1978pz}
%\bibitem{Kinnersley:1978pz}
 % W.~Kinnersley and D.~M.~Chitre,
  %``Symmetries of the Stationary Einstein-Maxwell Field Equations. 3.,''
  %J.\ Math.\ Phys.\  {\bf 19}, 1926 (1978).
  %%CITATION = JMAPA,19,1926;%%
  %\cite{Mazur:1983vi}



  [hep-th/9909099].
%\cite{Breitenlohner:1987dg}
\bibitem{Breitenlohner:1987dg}
  P.~Breitenlohner, D.~Maison, G.~W.~Gibbons,
  %``Four-Dimensional Black Holes from Kaluza-Klein Theories,''
  Commun.\ Math.\ Phys.\  {\bf 120}, 295 (1988).

\bibitem{Maison:2000fj}
  D.~Maison,
  %``Duality and hidden symmetries in gravitational theories,''
  Lect.\ Notes Phys.\  {\bf 540}, 273-323 (2000).

%\cite{Clement:2008qx}
\bibitem{Clement:2008qx}
  G.~Cl\'ement,
  ``Sigma-model approaches to exact solutions in higher-dimensional gravity and
  supergravity'',
arXiv:0811.0691 [hep-th].

%\cite{Galtsov:2008zz}
\bibitem{Galtsov:2008zz}
  D.~V.~Gal'tsov,
  %``Generating solutions via sigma-models,''
  Prog.\ Theor.\ Phys.\ Suppl.\  {\bf 172}, 121 (2008)
  [arXiv:0901.0098 [gr-qc]].
  %%CITATION = ARXIV:0901.0098;%%

\bibitem{Mazur:1983vi}
  P.~O.~Mazur,
  %``A Relationship Between The Electrovacuum Ernst Equations And Nonlinear Sigma Model,''
  Acta Phys.\ Polon.\ B {\bf 14}, 219 (1983).
  %%CITATION = APPOA,B14,219;%%

\bibitem{TYM} S. Tomizawa, Y. Yasuii and Y. Morisawa, Class. Quantum
Grav. {\bf 26}, 145006 (2009) [arXiv:0809.2001 [hep-th]].

\bibitem{Compere09} G. Comp\`ere, S. de Buyl, E. Jamsin and A. Virmani,
Class. Quantum Grav. {\bf 26}, 125016 (2009) [arXiv:0903.1645
[hep-th]].

\bibitem{Compere10} G. Comp\`ere, S. de Buyl, S. Stotyn and A. Virmani,
JHEP 1011:133 (2010) [arXiv:1006.5464 [hep-th]].

\bibitem{5to3} G. Cl\'ement: J. Math. Phys. \textbf{49}, 042503
(2008), Erratum, J. Math. Phys. \textbf{49}, 079901 (2008)
[arXiv:0710.1192[gr-qc]]

%\cite{Hassan:1991mq}
\bibitem{Hassan:1991mq}
  S.~F.~Hassan and A.~Sen,
  %``Twisting classical solutions in heterotic string theory,''
  Nucl.\ Phys.\ B {\bf 375}, 103 (1992)
  [hep-th/9109038].
  %%CITATION = HEP-TH/9109038;%%
%\cite{Maharana:1992my}

\bibitem{Maharana:1992my}
  J.~Maharana and J.~H.~Schwarz,
  %``Noncompact symmetries in string theory,''
  Nucl.\ Phys.\ B {\bf 390}, 3 (1993)
  [hep-th/9207016].
  %%CITATION = HEP-TH/9207016;%%
%\cite{Sen:1994fa}

\bibitem{Sen:1994fa}
  A.~Sen,
  %``Strong - weak coupling duality in four-dimensional string theory,''
  Int.\ J.\ Mod.\ Phys.\ A {\bf 9}, 3707 (1994)
  [hep-th/9402002].
  %%CITATION = HEP-TH/9402002;%%

\bibitem{sorokin}
%\bibitem{Pasti:1996vs}
  P.~Pasti, D.~P.~Sorokin and M.~Tonin,
  %``On Lorentz invariant actions for chiral p forms,''
  Phys.\ Rev.\ D {\bf 55}, 6292 (1997)
  [hep-th/9611100];
  %%CITATION = HEP-TH/9611100;%%
%\bibitem{Dall'Agata:1997db}
  G.~Dall'Agata, K.~Lechner and M.~Tonin,
  %``Covariant actions for N=1, D = 6 supergravity theories with chiral bosons,''
  Nucl.\ Phys.\ B {\bf 512} (1998) 179
  [hep-th/9710127].
  %%CITATION = HEP-TH/9710127;%%
  %\cite{Sen:1994wr}

\bibitem{Sen:1994wr}
  A.~Sen,
  %``Strong - weak coupling duality in three-dimensional string theory,''
  Nucl.\ Phys.\ B {\bf 434}, 179 (1995)
  [hep-th/9408083].
  %%CITATION = HEP-TH/9408083;%%

%\cite{Maison:1979kx}
\bibitem{Maison:1979kx}
  D.~Maison,
  %``Ehlers-harrison Type Transformations For Jordan's Extended Theory Of Gravitation,''
  Gen.\ Rel.\ Grav.\  {\bf 10}, 717-723 (1979).

\bibitem{ErnstWild76} F.J. Ernst and W.J. Wild, J. Math. Phys. {\bf 17},
182 (1976).

\bibitem{oxi} G.~Cl\'ement and D.~V.~Gal'tsov,
%``Oxidation of $D=3$ cosets and Bonnor dualities in $D\le6$'',
Phys.\ Rev.\ D {\bf  },  (2013)

\bibitem{gilmore}
  R.~Gilmore,
  {\it Lie Groups, Lie Algebras and Some of Their Applications},
%  Kieger Pub. Co (1974)
  John Wiley \& Sons, 1974.

\end{thebibliography}
\end{document}